\documentclass[showpacs,superscriptaddress,floatfix,amsmath,amssymb,twocolumn,pra,aps,longbibliography]{revtex4-1}
\usepackage{bm}
\usepackage[dvips]{graphicx}
\usepackage{wrapfig,subfigure}
\usepackage{stmaryrd}
\usepackage{color}
\usepackage[normalem]{ulem}
\usepackage{enumerate}
\usepackage{epstopdf}

\def\gb{{\bf g}}

\def\HR{H_{\rm RWA}}

\def\ep{\varepsilon}

\newcommand{\Ket}[1]{\vert #1 \rangle}

\def\exctd{\vert\tilde 1\rangle}
\def\ground{\vert \tilde 0\rangle}

\begin{document}
\author{M. I. Dykman}
\affiliation{Department of Physics and Astronomy, Michigan State
  University, East Lansing, Michigan 48824, USA}

\title{Coherent multiple-period states of periodically modulated qubits}

\date{\today}
\begin{abstract}
We consider multiple-period states in systems of periodically modulated qubits. In such states the discrete time-translation symmetry imposed by the modulation is broken. We explicitly show how multiple-period states emerge in the simplest quantum system, a single qubit subjected to a pulsed resonant modulation and/or a pulsed modulation of the transition frequency.  We also show that a qubit chain with the qubit coupling modulated at twice the qubit frequency has symmetry that allows mapping it on the Kitaev chain and thus provides an example of a topologically nontrivial Floquet system. An explicit solution for a two-qubit system illustrates the effect of  resonant period doubling for coupled qubits, whereas in a long chain period doubling is topologically protected.
\end{abstract}
\maketitle

\section{Introduction}

Breaking of the discrete time-translation symmetry (DTTS) is well known in classical physics. The simplest example is a parametric oscillator, which vibrates at half the frequency of the field that modulates the oscillator eigenfrequency \cite{Landau2004a}. 
Another example is the period doubling route to chaos in nonlinear dynamical systems, including periodically modulated systems \cite{Feigenbaum1978,*Feigenbaum1979,Huberman1979,Linsay1981}. DTTS breaking in driven classical systems due to many-body effects, including phase-transition like features of the onset of such symmetry breaking, has been also known \cite{Goldstein2018,Kim2006,Heo2010}. 

Recently much attention has attracted DDTS breaking in driven many-body quantum systems, sometimes called the time-crystal effect, see \cite{Khemani2016a,Else2016,Zhang2017,Choi2017,Pal2018,Rovny2018,Berdanier2018,Sacha2018,Dykman2018,O'Sullivan2018} and papers cited therein. In principle, the occurrence of subharmonics in a quantum multiple-state system is not surprising. Indeed, suppose one projects all Floquet eigenvalues (quasienergies) $\ep_n$ onto the first Brillouin zone $-\hbar\omega_F/2\leq \ep_n<\hbar\omega_F/2$, where $n$ enumerates the states and $\omega_F$ is the drive frequency. For a large number of states, the band will be filled almost densely \cite{Hone1997,Kohn2001}. One can then find a pair of states $\Ket{n_1}$ and $\Ket{n_2}$ with the quasienergy difference $\ep_{n_1}-\ep_{n_2}\approx  \hbar\omega_F/N$ with an integer $N>1$. If the system is in a superposition of these states, the expectation values of the dynamical variables oscillate  at the frequency $\omega_F/N$. However, the amplitude of such oscillations is determined by the overlap integral of  $\Ket{n_1}$ and $\Ket{n_2}$ and is often exponentially small. In addition, preparing the corresponding superposition is not necessarily trivial. 

With the rapid progress in making highly coherent and well-controlled systems of qubits, a natural question to ask is: How simple is it to achieve time-translation symmetry breaking in a quantum system? One of the goals of this paper is to show explicitly that a multiple-period superposition of states with an arbitrary multiplicity can emerge already in the simplest coherent quantum system, a periodically driven qubit. Achieving this goal requires finding an appropriate driving protocol that would be easy to implement in the experiment. The argument of the previous paragraph regarding almost dense quasienergy spectrum does not apply to a qubit, as it has only two states. On the other hand, by construction the overlap integral of the corresponding states is of order unity, which should facilitate an observation of the symmetry breaking. 

Another problem addressed in the paper is a qubit chain with the modulated coupling between the qubits.  We consider modulation close to twice the qubit eigenfrequencies. Looking for time-symmetry breaking in such a system is motivated by the symmetry breaking that occurs in a parametrically modulated classical oscillator. Unexpectedly, we find that the dynamics of a modulated qubit chain maps onto that of the Kitaev chain. This is a consequence of the symmetry of the modulated spin system. A solution that would explicitly demonstrate the underlying symmetry should be sought first for two coupled qubits. For a long chain, on the other hand, we seek to establish the relation between the familiar onset of Majorana fermions in the topologically nontrivial regime and the onset of period doubling. 

Our analysis refers to the regime where the transition frequency of the qubit(s) $\omega_0$ is large compared to the Rabi frequency, for a single qubit,  and compared to the qubit coupling in the frequency units, for the qubit chain. This suggests separating the time scales and analyzing the dynamics in slow time compared to $2\pi/\omega_0$, i.e., in the rotating frame. However, the quasienergy is defined in the laboratory frame. Therefore to describe the time-symmetry breaking, the results of the analysis have to be projected back to this frame.

\section{A periodically driven qubit}

We start with a single periodically modulated qubit. The modulation includes a resonant drive with a modulated amplitude and a low-frequency modulation of the interlevel spacing of the qubit. The Hamiltonian in the laboratory frame is
\begin{align}
\label{eq:H_lab_frame}
H_{\rm lab} = \frac{1}{2}[\omega_0+ \nu(t)] \sigma^z +F(t)\sigma^x \cos\omega_F t , 
\end{align}
with $|\omega_F-\omega_0|\ll \omega_F$. We have set $\hbar = 1$; $\sigma^{x,y,z}$ are the Pauli matrices.  The wave functions and the quasienergy spectrum of the spin with a time-dependent Hamiltonian have been found in several limiting cases, in particular, for a sinusoidal modulation [$F(t)=$const, $\nu(t)=0$] \cite{Ma2007},
see also 
\cite{Schmidt2018a,Schmidt2018} 
and references therein.

We assume that $\nu(t) $ and $F(t)$ have the form of periodic pulses,
\begin{align}
\label{eq:modulation_form}
&\nu(t) = \nu_1\sum_n\overline\delta (t-nT), \nonumber\\
& F(t) = F_0+ F_1\sum_n\overline\delta(t-nT).
\end{align}
Here, the period $T$ is a multiple of the drive period, $T=2\pi n/\omega_F$ with integer $ n \gg 1$; in this case the Hamiltonian $H_{\rm lab}$ as a whole is periodic with period $T$. We use the notation $\overline\delta(t)$ for a function, which is smooth on the time scale $\sim 1/\omega_F$, but looks like a $\delta$-function on the scale $\sim \min(T, |\omega_F-\omega_0|^{-1})$ and has unit area, $\int_{-\infty}^\infty dt \overline\delta (t)=1$. The pulses can be applied  independently to the level spacing and to the driving force amplitude, but when the both parameters are modulated, we assume that they are modulated with the same period. The pulse strengths $\nu_1, F_1$ are dimensionless (to incorporate $\hbar$, we have to replace $\nu_1\to \hbar\nu_1, F_1\to \hbar F_1$).

If the modulation is comparatively weak, so that $|\nu(t)|,\,|F(t)|\ll \omega_F$, one can go to the rotating frame using the transformation $U(t)=\exp(-i\omega_Ft\sigma_z/2)$. In the rotating wave approximation (RWA) the Hamiltonian becomes
\begin{align}
\label{eq:H_RWA}
&H_{\rm RWA} =  \frac{1}{2}\Delta \sigma^z +\frac{1}{2}F_0\sigma^x + \frac{1}{2}{\bf g}{\bm \sigma}\sum_n \delta(t-nT), \nonumber\\
&\Delta = \omega_0 -\omega_F, \qquad g^x = F_1, \quad g^z=\nu_1, \quad g^y=0 .
\end{align}
Here we have taken into account that, in slow time, the function $\overline\delta(t)$ becomes a $\delta$-function. The components of the vector ${\bf g}$ are determined by the modulation strength. We emphasize that the Hamiltonian $\HR$ depends on time and, as $H_{\rm lab}$, is periodic with period $T$.

The time-independent part of $\HR$ describes a spin in an effective magnetic field with $z$ and $x$ components being $\Delta$ and $F_0$. It is convenient to rotate the qubit in such a way that the new $z$-axis is pointing along this field, $\sigma^z +i\sigma^x = \exp(i\phi)(\tilde\sigma^z +i\tilde\sigma^x)$. The rotation angle $\phi$ is given by the familiar equation  $\tan\phi=F_0/\Delta$. Then the RWA Hamiltonian becomes
\begin{align}
\label{eq:rotated_H_RWA}
H_{\rm RWA} =  \frac{1}{2}\Omega \tilde\sigma^z  + \frac{1}{2}\tilde{\bf g}\tilde{\bm \sigma}\sum_n \delta(t-nT), 
\end{align} 
where $\Omega=(\Delta^2 + F_0^2)^{1/2}$ is the Rabi frequency in the absence of the pulses and 
$\tilde g^z+i\tilde g^x = \exp(-i\phi)(g^z + i g^x)$, i.e., $\tilde\gb$ is the rotated vector ${\bf g}$. 

Prior to analyzing the dynamics in the rotating frame described by the RWA  Hamiltonian $\HR$, Eq.~(\ref{eq:rotated_H_RWA}), we note that the Floquet eigenstates of this time-periodic Hamiltonian  are also Floquet eigenstates of the Hamiltonian $H_{\rm lab}$ in the laboratory  frame. Indeed, if $\psi(t)$ is a Floquet eigenstate of $\HR$ with a rotating-frame quasienergy $\ep$, this means that 
\begin{align}
\label{eq:Floquet_condition}
i\dot \psi = \HR \psi, \qquad \psi(t+T) = \exp(-i\ep T)\psi(t).
\end{align}
Taking into account the form of the transformation to the rotating frame $U(t)$, we see that the corresponding wave function in the laboratory frame $\psi_{\rm lab}(t) = U(t) \psi(t)$ also satisfies the Floquet-eigenstate condition,  
\begin{align*}
\psi_{\rm lab}(t+T)& = U(t+T)\psi(t+T) =e^{-i\omega_FT\sigma_z/2}\,e^{-i\ep T}\psi_{\rm lab}(t)\\
&\equiv \exp (-i\ep T-i\pi n)\psi_{\rm lab}(t),
\end{align*}
where we used that $\omega_FT = 2\pi n$. The quasienergy $\ep$ in the rotating frame corresponds to the quasienergy $\ep +  n\pi T^{-1}$  in the laboratory frame projected on  the Brillouin zone of quasienergies $(-\pi/T,\pi/T)$. For an even $n$ such a projection gives $\ep$, whereas for an odd $n$ it gives $\ep-(\pi/T)\,{\rm sgn}\,\ep$, provided $-\pi/T\leq \ep< \pi/T$.  In what follows  we use the term ``rotating-frame quasienergy'' for the quasienergy defined by Eq.~(\ref{eq:Floquet_condition}).

\subsection{Finding the Floquet eigenstates in the rotating frame}

To find the Floquet wave functions we start by choosing an arbitrary instant $t_0$ within a period of $\HR$. We solve the time-dependent Schr\"odinger equation $i\dot \psi = \HR \psi$ with the initial condition $\psi(t_0)=\psi_0$ to find $\psi(t+t_0)$. We then find the explicit form of $\psi_0$ so that the solution meets the condition $\psi(T+t_0) = \exp(-i\ep T)\psi_0$. This gives both $\psi_0$ and the value of the rotating-frame quasienergy $\ep$. Clearly, $\psi_0$ depends on the chosen $t_0$, since the Floquet state is time-dependent, but $\ep$ does not. 

Since all periods are on equal footing, we can choose $t_0$ such that $0<t_0 <T$  [equivalently, we can choose $kT<t_0<(k+1)T$ with an arbitrary integer $k$]. A Floquet eigenstate at time $t_0$ can be written as a superposition 
\[\psi(t_0)\equiv \psi_0 =\alpha\exctd +\beta\ground,\] 
where $\exctd$ and $\ground$ are the eigenstates of $\tilde\sigma^z$ with eigenvalues $1$ and $-1$, respectively; these eigenstates form a complete set, which justifies the form of $\psi(t_0)$.  

The evolution of the wave function in the interval $(t_0,T-\epsilon)$ is controlled by the time-independent part of $\HR$; we formally consider the limit $\epsilon\to +0$; on physical grounds, $\epsilon\ll T, \Omega^{-1}$. At the time $T-\epsilon$ the wave function becomes 
\begin{align}
\label{eq:free_evolution}
\psi(T-\epsilon) =& \alpha \exp[-i\Omega(T-t_0)/2]\exctd\nonumber\\
& +\beta\exp[i\Omega(T-t_0)/2]\ground \quad (\epsilon\to +0).
\end{align}

During the pulse, the evolution of the wave function is controlled by the term $\propto \tilde\gb$, which is much larger than the time-independent part of $\HR$. The eigenfunctions of $\tilde\gb\tilde{\bm\sigma}$ can be chosen in the form
\begin{align}
\label{eq:g_eigenfunctions}
\psi_{\pm} = \zeta_{\pm}\exctd + \eta_{\pm}\ground, \quad \tilde\gb\tilde{\bm\sigma}\psi_{\pm} = \pm g\psi_{\pm},
\end{align}
where $g=|\gb|\equiv |\tilde\gb|$ and the real coefficients $\zeta_{\pm}, \eta_{\pm}$ are given by the equation
\begin{align}
\label{eq:theta_angle}
&\zeta_+ + i\zeta_- =\exp(-i\theta), \quad i\eta_+ +\eta_-=\exp(i\theta); \nonumber\\
& \tan \theta = \tilde g^x/(g+\tilde g^z).
\end{align}

The functions $\psi_{\pm}$ also form a complete set. Therefore one can write $\psi(T-\epsilon) = C_+\psi_+ + C_-\psi_-$ and express the coefficients $C_{\pm}$ as linear combinations of the coefficients $\alpha,\beta$ of the initial wave function $\psi_0$ with the weighting factors that depend on the parameters   $\theta$ and $\Omega(T-t_0)$, cf. Eqs.~(\ref{eq:free_evolution}) and (\ref{eq:g_eigenfunctions}). From Eqs.~(\ref{eq:rotated_H_RWA}) and (\ref{eq:g_eigenfunctions}) we see that, immediately after the pulse, the wave function is $\psi(T+\epsilon) = C_+\psi_+\exp(-ig/2) + C_-\psi_-\exp(ig/2)$. This wave function  can be again written in the basis of the functions $\exctd$ and $\ground$, $\psi(T+\epsilon) = \alpha_T\exctd +\beta_T\ground$.  The coefficients $\alpha_T, \beta_T$ are linear combinations of $C_{\pm}$ and ultimately are linear combinations of $\alpha, \beta$.

The evolution of $\psi(t)$ in the interval $(T+\epsilon, T+t_0)$ is again controlled by the Hamiltonian $\Omega\tilde\sigma^z/2$, with $\psi(T+t_0) = \alpha_T\exp(-i\Omega t_0/2)\exctd + \beta_T\exp(i\Omega t_0/2)\ground$. This expression relates $\psi(T+t_0)$ to $\psi(t_0)$. The condition (\ref{eq:Floquet_condition}) with $t=t_0$ now reads $\alpha_T\exp(i\Omega t_0/2)=\alpha\exp(-i\ep T), \; \beta_T\exp(-i\Omega t_0/2) = \beta\exp(-i\ep T)$. Given that $\alpha_T,\beta_T$ are linear combinations of $\alpha,\beta$, this gives a homogeneous system of linear equations for $\alpha,\beta$, which is  the eigenvalue problem for the quasienergy $\ep$. In fact, the corresponding characteristic equation is an equation for $\exp(-i\ep T)$. It has two solutions with equal $\cos(\ep T)$ and opposite in sign $\sin(\ep T)$.

The  values of the rotating-frame quasienergy $\ep_{1,2}$ obtained this way can be chosen in the form
\begin{align}
\label{eq:quasienergy}
&\ep_{1,2} = \pm T^{-1}\nonumber\\
&\times\arccos\left[\cos\frac{g}{2}\, \cos\frac{\Omega T}{2} -\cos 2\theta\,\sin\frac{g}{2}\, \sin\frac{\Omega T}{2}\right].\end{align} 
They correspond to two Floquet eigenstates of the qubit in the rotating frame, $\psi_{j}(t+T)=\exp(-i\ep_jT)\psi_{j}(t)$ ($j=1,2$), and thus to two Floquet eigenstates in the laboratory frame.

\begin{figure}[h]
\includegraphics[scale=0.45]{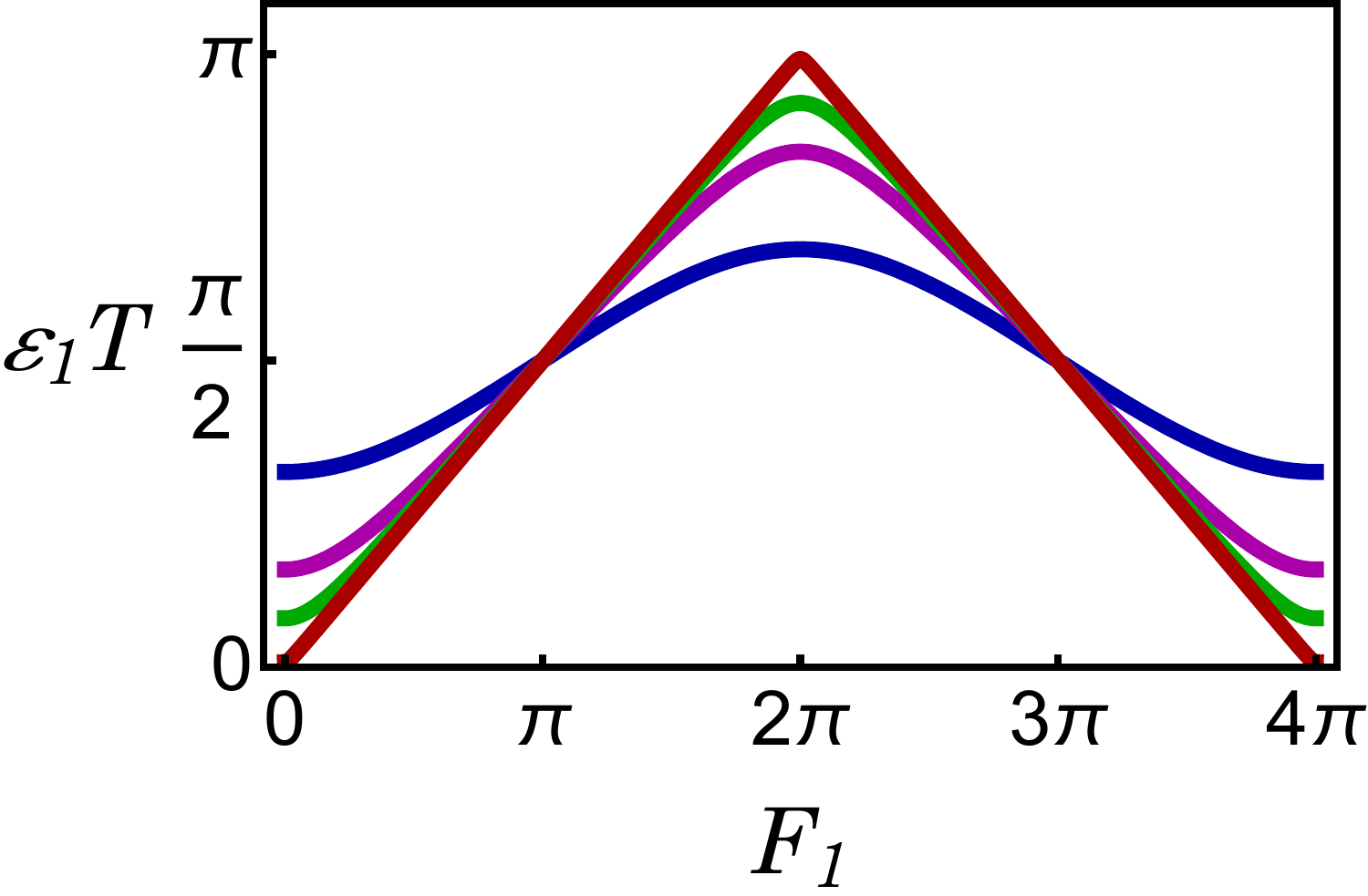}
\caption{The quasienergy $\ep_1$ of a qubit in the rotating frame. The qubit is driven at its eigenfrequency, $\omega_F = \omega_0$. The driving amplitude is pulsed with period $T \gg 2\pi/\omega_F$,  the dimensionless area of a pulse is $F_1$.  The blue, magenta, green, and red curves (the curves from bottom to top, for $F_1=2\pi$) refer to the Rabi frequency $\Omega = 2T^{-1}, T^{-1},0.5T^{-1}$, and  $0.05T^{-1}$.}
\label{fig:single_qubit}
\end{figure}

In Fig.~\ref{fig:single_qubit}  we show the quasienergy in the case where only the amplitude of the resonant field $F(t)$ is pulsed, whereas the frequency of the qubit is not modulated, $\nu_1=0$.  As seen from the figure and also from Eq.~(\ref{eq:quasienergy}), the quasienergy is a periodic function of the intensity (area) of the pulses $F_1 = g$ with period $4\pi$. It is also a periodic function of the Rabi frequency $\Omega$ with period $4\pi/T$. The magnitude of the oscillations of $\ep_1$ with $F_1$ is maximal for $\Omega T = n\pi$ with an integer $n$.

A linear combination $\Psi(t) = A\psi_1(t) + B\psi_2(t)$ is not a Floquet eigenstate, $\Psi(t+T)\neq \exp(-i\ep T)\Psi(t)$ with a real $\ep$ for $AB\neq 0$. However, if $\ep_1-\ep_2 = 2\ep_1 = 2\pi M/NT$ with integer  $M,N$ and $N>|M|\geq 1$, we have $\Psi(t+NT) = \exp(-iM\pi)\Psi(t)$. Physical observables in such a state are oscillating with period $NT$. The amplitude of the oscillations is $\propto |AB|$ and therefore can be of order one. Another single-particle system where the difference of the quasienergies can be made equal to $2\pi/NT$ (with $N=2$ and $3$) while keeping the overlap of the corresponding wave functions $\sim 1$ is a resonantly modulated nonlinear oscillator \cite{Marthaler2007,Zhang2017b,Zhang2017c}.

\subsection{Observing multiple-period states of a qubit}

A particularly simple way to obtain multiple-period states is to pulse the level spacing of a qubit, $\omega_0\to \omega_0+\nu(t)$, with $\nu(t)$ of the form given by Eq.~(\ref{eq:modulation_form}). The periodic pulsing leads to an increment of the phase difference between the states $|1\rangle$ and $|0\rangle$ by $\nu_1$ each period $T$. It is obvious without calculation that, for $\omega_0=2\pi n/T$ with an integer $n$, the states $|1\rangle$ and $|0\rangle$ are quasienergy states in the laboratory frame with the quasienergies $\pm [\omega_0 /2+ (\nu_1/2T)]$ projected onto the Brillouin zone $(-\pi/T, \pi/T)$. Because of the relation between the quasienergies in the laboratory and rotating frames, this expression  coincides with what follows from Eq.~(\ref{eq:quasienergy}). To see this one should set in Eq.~(\ref{eq:quasienergy})  $\Omega = \theta =\phi = 0, g=\nu_1$, which then gives $\ep_{1,2}=\pm T^{-1}\arccos[\cos( \nu_1/2)]$. 

The pulsing-induced phase shift can be revealed in a standard way using Ramsey fringes. The measurement is done in the following way. If the  qubit with a pulsed level spacing is in the state $|0\rangle$, one applies a $\pi/2$ resonant (at frequency $\omega_0$) pulse  at some time $t=t_i$. This pulse  drives the qubit into a superposition of the quasienergy states with equal weights. If at a time $t_f$ there is applied another resonant $\pi/2$ pulse at frequency $\omega_0$, the population of the excited state, which is given by the projection of the wave function after the pulse $\Psi(t_f)$ on the wave function of the excited state $|1\rangle$, becomes 
\[|\langle 1|\Psi(t_f)\rangle|^2 = \cos^2\Bigl[ \nu_1 \Bigl\lfloor (t_f - t_i)/T)\Bigr\rfloor /2\Bigr],\]
where, $\lfloor x\rfloor$ is the integer part of $x$. For $\nu_1 = 2\pi M/N$ this population is a periodic function of $t_f-t_i$ with period $NT$.

Another example is where the qubit is driven only by  a close to resonance field $F(t)\cos\omega_Ft$ with a pulsed envelope, $F(t)=F_1\sum_n\overline\delta(t-nT)$. In particular, for exact resonance, $\omega_F = \omega_0$, we have in Eq.~(\ref{eq:quasienergy})  $\phi=\Omega=0, \theta=\pi/4, g=F_1$, and then $\ep_{1,2}=\pm T^{-1}\arccos[\cos(F_1/2)]$. The quasienergy wave functions in the rotating frame are $\psi_{\pm} = (|0\rangle \pm |1\rangle)/\sqrt{2}$. If  in the laboratory frame the qubit is initially ($t=t_i$) in the ground state $|0\rangle_{\rm lab}$, the occupation of the excited state $|1\rangle_{\rm lab}\equiv U(t)|1\rangle$ at time $t_f$ is
\[|\langle 1|\Psi(t_f)\rangle|^2 = \sin^2\Bigl[ F_1 \Bigl\lfloor (t_f - t_i)/T)\Bigr\rfloor /2\Bigr].\]
If the pulse area is a simple fraction of $2\pi$, $F_1=2\pi M/N$, the state population varies in time periodically with period $NT$. 

In the both examples, establishing the broken DTTS is reduced to the standard operation of detecting the occupation of the excited state of a qubit. We emphasize again that the pulses  must be smooth on the scale  $\omega_0^{-1}$ but short on the scale $T$.

\section{A qubit chain with a resonantly modulated coupling}

An interesting behavior, including the possibility of DTTS breaking, is provided by a system of coupled qubits with a resonantly modulated coupling. We consider here a sinusoidal modulation at frequency $\omega_F$ close to twice the qubit transition frequency $\omega_0$, i.e., $|\omega_F-2\omega_0|\ll \omega_0$. The Hamiltonian of a chain of coupled qubits with nearest neighbor coupling and with one of the coupling parameters, $J_{xx}$, being modulated has the form
\begin{align}
\label{eq:chain_lab}
&H_{\rm lab}^{\rm (chain)}=\frac{1}{2}\omega_0 \sum_n \sigma_n^z -J_{xx}(t)\sum_n\sigma_n^x\sigma_{n+1}^x\nonumber\\
& - J_{yy}\sum_n\sigma_n^y\sigma_{n+1}^y, 
\qquad J_{xx}=J_{xx}^{(0)} + 2F\cos\omega_Ft.
\end{align} 
Controllable $xx$ coupling has been implemented in several types of flux qubits, cf. \cite{Yan2018} and references therein, although we are not aware of the experiments where the coupling was modulated at frequency $\sim 2\omega_0$. However, we do not immediately see physical constraints that would prohibit such a modulation, although its implementation could be accompanied by a decrease of the coherence time. The results below apply also to the case where the both coupling parameters, $J_{xx}$ and $J_{yy}$, are modulated.

In many implementations, the qubit coupling is weak compared to the qubit transition energy, $|J_{xx}|, |J_{yy}|\ll \omega_0$, which we assume to be the case. Then, as shown below, resonant driving can lead to strong effects even where it is comparatively weak, $|F|\ll \omega_0$.  For small $|J_{xx}|, |J_{yy}|, |F|$ the dynamics can be conveniently analyzed by switching to the rotating frame at frequency $\omega_F/2$. The transformation is 
\[U^{\rm (chain)}(t)=\exp[-i(\omega_F t/4)\sum_n(\sigma_n^z+I_n)],\]
where $I_n$ is the identity operator for an $n$th spin, i.e., the $2\times 2$ unit matrix operating in the $n$th spin Hilbert space; as will be seen below, particularly when the problem is formulated in terms of the fermion operators, introducing the operators $I_n$ into $U^{\rm (chain)}(t)$ simplifies the analysis.  

In the rotating wave approximation the Hamiltonian becomes 
\begin{align}
\label{eq:chain_RWA}
&H_{\rm RWA}^{\rm (chain)} = -\frac{1}{2}\mu \sum_n \sigma_n^z -\frac{1}{4}J\sum_n(\sigma_n^+\sigma_{n+1}^- +  \sigma_{n+1}^+\sigma_{n}^-)\nonumber\\
& -
\frac{1}{4}F\sum_n(\sigma_n^+\sigma_{n+1}^+ + \sigma_{n+1}^-\sigma_{n}^-), \quad \mu=\frac{1}{2}\omega_F-\omega_0.
\end{align}
Here $J= J_{xx}^{(0)} + J_{yy}$ and $\sigma_n^\pm = \sigma_n^x \pm i\sigma_n^y$. In the rotating frame, the frequency detuning $\mu$ plays the role of the scaled magnetic field along the $z$-axis. The detuning is small compared to $\omega_0$, but can be of the same order of magnitude as the coupling parameter $J$ and the modulation amplitude $F$. In the expression for $\HR^{\rm (chain)}$ we left out the identity operator in the spin-chain Hilbert space $-(\omega_F/4)\sum_nI_n$, as it does not affect the dynamics. 

The Hamiltonian $\HR^{\rm (chain)}$ is independent of time. Therefore, in contrast to the previously considered case of the pulse-modulated single qubit,  the eigenvalues and eigenfunctions in the rotating frame in the RWA are given by the solution of the stationary problem 
\begin{align*}
\HR^{\rm (chain)}\psi = \ep \psi.
\end{align*}
The rotating-frame eigenfunction $\psi$ with an RWA eigenvalue $\ep$ evolves in time as
\begin{align}
\label{eq:psi_chain_vs_time}
\psi(t+T)=e^{-i\ep T}\psi(t). 
\end{align}
This equation holds for an arbitrary $T$, but in what follows we will be interested in $T$ being the modulation period,  $T=2\pi/\omega_F$.

To relate $\ep$ to the quasienergy in the laboratory frame, we introduce the parity operator $P$,
\begin{align}
\label{eq:parity}
P=\exp\Bigl[-i\frac{\pi}{2}\sum_n(\sigma_n^z+I_n)\Bigr], \quad [P,\HR^{\rm(chain)}]=0.
\end{align}
Clearly, $P=P^\dagger$ and $P^2 = \prod_nI_n$. The eigenvalues of $P$ are $\pm 1$. The parity operator $P$ commutes not only with $\HR^{\rm(chain)}$, but also with the  Hamiltonian in the laboratory frame, $[P, H_{\rm lab}^{\rm (chain)}]=0$. The parity conservation is not a consequence of the RWA. 

%
%
%
%

With the account taken of the explicit form of the unitary transformation $U^{\rm(chain)}(t)$ and Eq.~(\ref{eq:psi_chain_vs_time}), we see that the wave function $\psi_{\rm lab}(T)=U^{\rm (chain)}(t)\psi(t)$ in the laboratory frame transforms over the period as 
\begin{align*}
\psi_{\rm lab}(t+T) =&\exp\Bigl[-i(\omega_F T/4)\sum_n(\sigma_n^z+I_n)\Bigr]e^{-i\ep T}\psi_{\rm lab}(t)\\
&= \exp(-i\ep T)P\psi_{\rm lab}(t).
\end{align*}
Therefore for even-parity states, i.e., where the eigenvalue of $P$ is $1$, the rotating-frame quasienergy $\ep$ is also the quasienergy in the laboratory frame, whereas for odd-parity  states, where the eigenvalue of $P$ is $-1$, the quasienergy in the laboratory frame is $\ep - (\omega_F/2)\,{\rm sgn}\,\ep$, i.e., it is shifted from $\ep$ by half the Brillouin zone. 

Importantly, the eigenstates of $\HR^{\rm (chain)}$ with different parity can have the same eigenvalue $\ep$, there is no level repulsion between such states. From this general argument, it is clear that, if we prepare the system in a combination of states with the same $\ep$ in the rotating frame, but with different parity, so that in the laboratory frame the wave function is $U^{\rm(chain)}(t)[\alpha_{\rm even}\psi_{\rm even}(t) + \alpha_{\rm odd}\psi_{\rm odd}(t)]$, the expectation values of dynamical variables in this state will have period $2T$. In other words, this state will have broken time-translation symmetry.

\subsection{A two-qubit chain}

To illustrate the occurrence of the breaking of the time-translation symmetry in a modulated chain we start with a particularly simple case of two qubits, i.e., we assume that $n=1,2$ in Eq.~(\ref{eq:chain_RWA}). The eigenfunctions and eigenvalues of the operator $\HR^{\rm (chain)}$ in this case are
\begin{align}
\label{eq:two_qubit}
&\psi_{1,2} = \cos\phi_{1,2}\Ket{00} + \sin\phi_{1,2}\Ket{11}, \quad \nonumber\\
&\psi_{3,4}= (\Ket{01}\pm \Ket{10})/\sqrt{2},\nonumber\\
&\ep_{1,2}=\pm (\mu^2+F^2)^{1/2},\quad  \ep_{3,4} = \mp J,
\end{align}
where $\tan \phi_{1,2}=(\mu-\ep_{1,2})/F$. The eigenvalue of the parity operator on the functions $\psi_{1,2}$ is 1, whereas on the functions $\psi_{3,4}$ it is $-1$. 

\begin{figure}[h]
\includegraphics[scale=0.3]{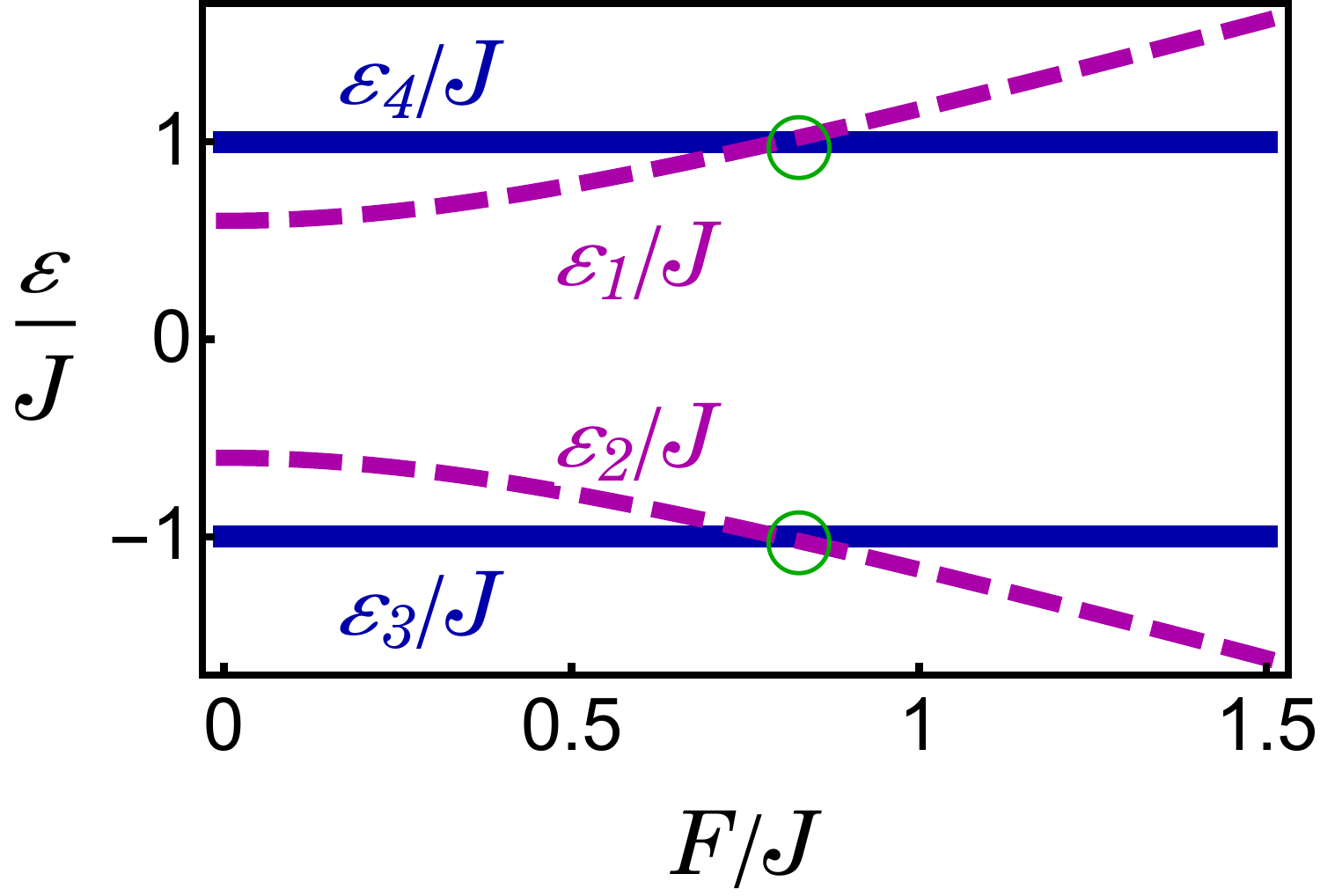}
\caption{The rotating-frame energies of a two-qubit system with the coupling periodically modulated at frequency $\omega_F$ close to twice the qubit transition frequency $\omega_0$, Eq. (\ref{eq:two_qubit}). The dashed and solid curves refer to the states 1, 2 and 3, 4, respectively. The frequency detuning is $\frac{1}{2}\omega_F - \omega_0 \equiv \mu =0.6J$. The green circles indicate where the quasienergies in the laboratory frame differ by $\omega_F/2$. There is no anticrossing between the corresponding quasienergy levels.}
\label{fig:quasienergies}
\end{figure}

The  quasienergies in the rotating frame, Eq.~(\ref{eq:two_qubit}), as functions of the driving amplitude $F$ are shown in Fig.~\ref{fig:quasienergies}. Only the values of $\ep_{1,2}$, which refer to the even-parity states,  vary with $F$. If $|\mu| <|J|$ (for concreteness, we assume $J>0$), by varying $F$ one can make the rotating-frame quasienergies of the states with different parity coincide, so that $\ep_1 = \ep_4$ and  $ \ep_2 = \ep_3$. In this case, as explained above, linear combinations $\alpha_{\rm even}\psi_1 + \alpha_{\rm odd} \psi_4$ and $\alpha_{\rm even}'\psi_2 + \alpha_{\rm odd}' \psi_3$ are period-two states in the laboratory frame. 

The analysis immediately extends to the case where the qubit frequencies are slightly different. The parity is still conserved in this case. One can easily check that the rotating-frame quasienergies of the different-parity states become equal simultaneously for two pairs of states, as in the case of equal qubit frequencies.

To observe the period doubling one can prepare the system in a superposition of states contained in $\psi_{1,2}$ and $\psi_{3,4}$, for example, in a superposition of states $|01\rangle$ and $|00\rangle$. Then if the driving is tuned so that $(\mu^2+F^2)^{1/2}  = |J|$, the expectation values of physical observables will oscillate with period $2T$, the period doubling effect.

\subsection{Topologically nontrivial Floquet regime}

In the case of a longer modulated qubit chain, the analysis of the dynamics can be conveniently done using the Jordan-Wigner transformation from spins to spinless fermions. If the creation and annihilation operators of a fermion on site $n$ are $a_n^\dagger$ and $a_n$, respectively, the Hamiltonian (\ref{eq:chain_RWA}) becomes the Hamiltonian of the Kitaev chain 
\begin{align}
\label{eq:Kitaev}
H_K &=  -\mu\sum_n\Bigl(a_n^\dagger a_n -\frac{1}{2}\Bigr)- J\sum_n(a_n^\dagger a_{n+1} +a_{n+1}^\dagger a_n)\nonumber\\
& -F\sum_n(a_n^\dagger a_{n+1}^\dagger +a_{n+1}a_n).
\end{align}
In the fermion representation in the rotating frame, the role of the chemical potential $\mu$ is played by the frequency detuning $(\omega_F/2)-\omega_0$, cf. Eq.~(\ref{eq:chain_RWA}).

In the fermion representation, the parity operator (\ref{eq:parity}) takes a simple form
\begin{align}
\label{eq:parity_fermion}
P=\exp\Bigl(-i\pi\sum_n a_n^\dagger a_n\Bigr), \quad [P, H_K]=0.
\end{align}
The eigenvalues of $P$ are $-1$ and $1$ for odd and even number of fermions, respectively. 

The properties of the Kitaev chain are well-known  \cite{Kitaev2001,Alicea2012}. The considered Floquet system is topologically nontrivial for $|\mu|\equiv |\frac{1}{2}\omega_F - \omega_0|< 2|J|$. Interestingly, this condition on the detuning of the drive frequency or the strength of the qubit coupling is less restrictive than the condition $|\mu|<|J|$ that must be met, as seen from Eq.~(\ref{eq:two_qubit}), to obtain period-two states in a system of two qubits. Both conditions can be met for a given qubit system by tuning the drive frequency closer to $2\omega_0$. 

\begin{figure}[h]
\includegraphics[scale=0.265]{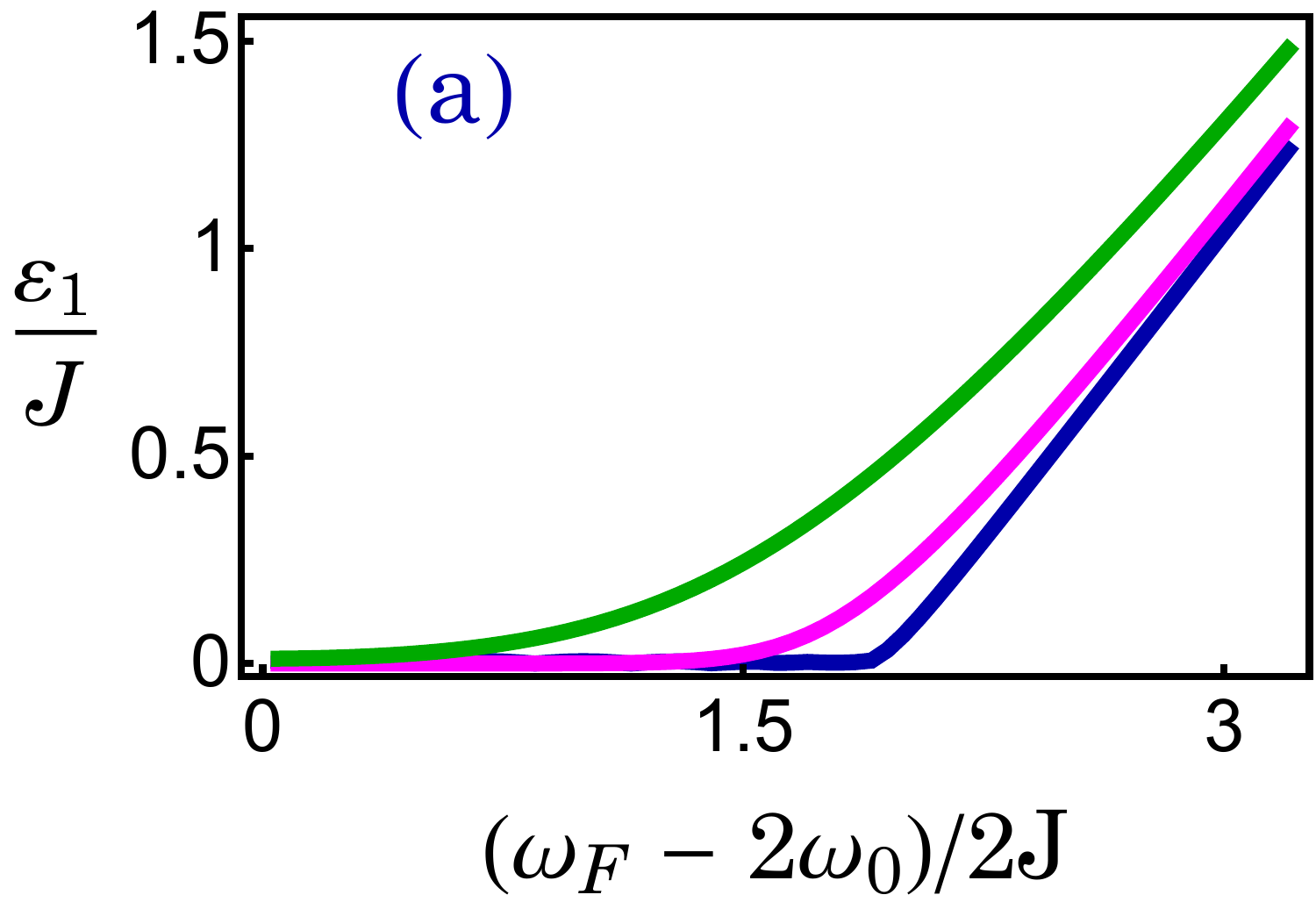} \hfill
\includegraphics[scale=0.26]{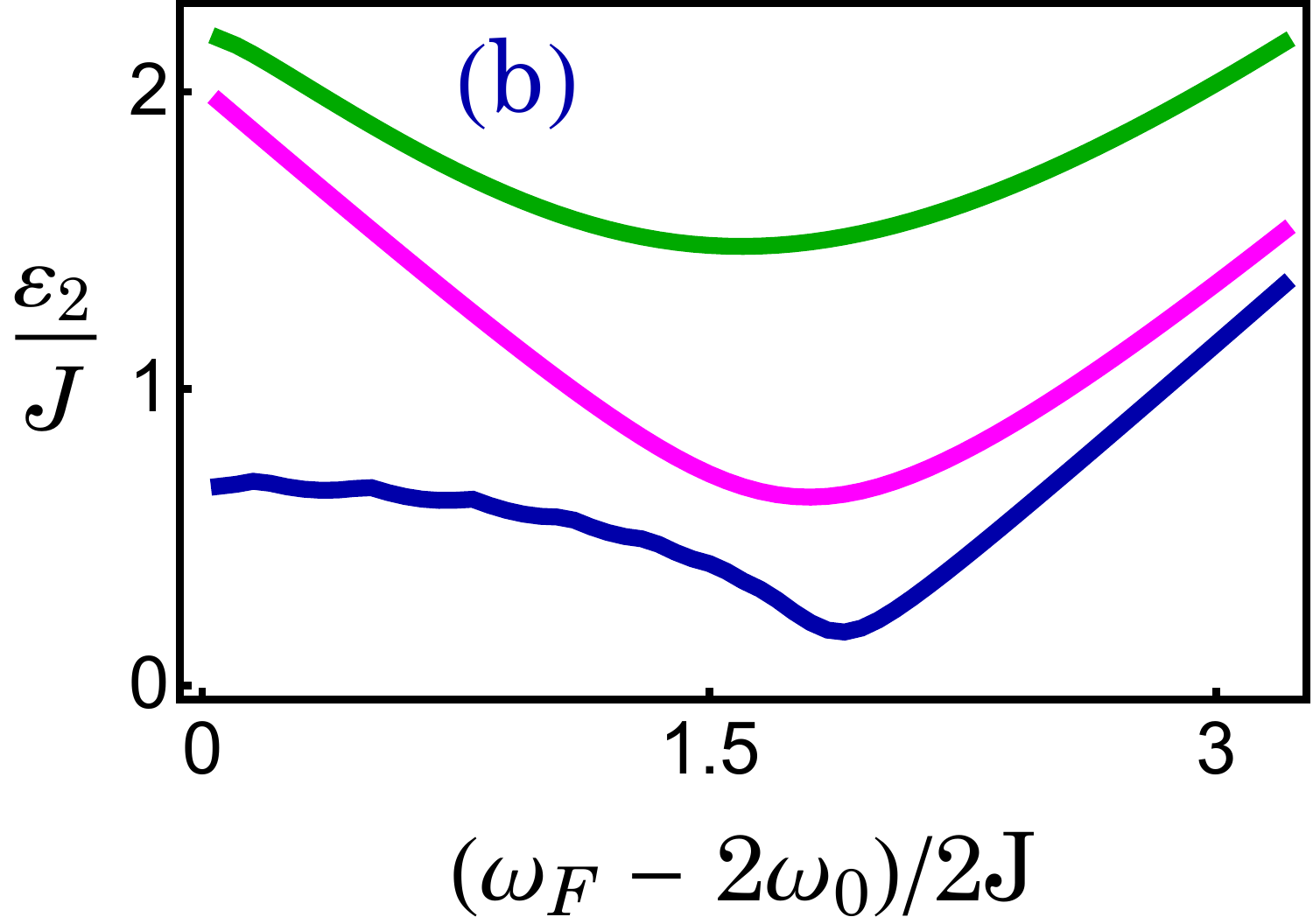} 
\caption{The rotating-frame energies of the first [panel (a)] and second [panel (b)] lowest excited states of the qubit chain with a resonantly modulated coupling. The energies are given by the eigenvalues of the Hamiltonian (\ref{eq:Kitaev}) for a 16-sites chain of fermions. The chemical potential of the fermions $\mu$ is given by the frequency detuning  $(\omega_F-2\omega_0)/2$, see Eq.~(\ref{eq:chain_RWA}). The blue, magenta, and green curves (the lowest to the highest curves) correspond to the scaled modulation amplitude $F/J = 0.3, 1.2$, and $3$.
}
\label{fig:Kitaev}
\end{figure}

For completeness, the familiar behavior of excitations in the Kitaev chain is illustrated in Fig.~\ref{fig:Kitaev}  for a comparatively short modulated qubit array. The data are obtained using the method \cite{Lieb1961} and show the evolution of the first and second excited states in the rotating frame with the changing modulation frequency $\omega_F$. The lowest excited state for small $|\mu/J|$ corresponds to the excitation localized at the edges of the chain and is described by the Majorana physics \cite{Kitaev2001}. As expected, its energy becomes extremely small  for small $|\mu/J|$.  For example, for the chain of 16 qubits in Fig.~\ref{fig:Kitaev}, $\ep_1/J <4\times 10^{-5}$ for $|\mu/J|<1$ and $F/J=0.3$, and  $\ep_1/J <2\times 10^{-7}$ for $|\mu/J|<1$ and $F/J=1.2$; for small $F/J$ the boundary effects make $\ep_1/J$ oscillate with $|\mu/J|$ with a very small amplitude. From the point of view of obtaining small $\ep_1$ for $|\mu/J|\sim 1$, the optimal range of the driving amplitude is where $F/J$ is close to 1; this range depends on the chain length.

The considered implementation of the Kitaev chain is based on resonant modulation of a qubit system and is thus qualitatively different from the proposed implementations that use flux qubits or other Josephson-junction based qubit systems with no time-periodic modulation, see \cite{Narozhny2017,Backens2017} and references therein. It is also different from the  superconducting cold-atom quantum wire where, for  a periodically step-wise modulated chemical potential, there emerge two pairs of localized Majorana fermions  with the quasienergy difference equal to $\pi/T$, with  $T$ being the modulation period \cite{Jiang2011}. Bound states with 0 and $\pi$ effective quasienergies were found also in a linear optical system that mimics a periodically modulated system \cite{Kitagawa2012}. Not only our analysis refers to a system of a different type, but as we have shown, resonant modulation of a qubit chain essentially automatically ``builds in'' a pair of Majorana fermions with the quasienergy shifted from the ground state by $\omega_F/2$. 

Implementing the Floquet-Kitaev chain (\ref{eq:Kitaev}) with qubits is advantageous in terms of simulating Majorana fermions and obtaining a topologically protected period-2 state in a controlled way. The ground state of the spin chain in the absence of the driving corresponds to all spins aligned along the $z$-axis, with the expectation value of $\sigma_n^z$ being $-1$ for all $n$. It corresponds to the vacuum state of the fermions, $\langle a_n^\dagger a_n\rangle =0$ for all $n$. The lowest excited state of the chain without driving has an opposite parity. In a finite-length chain the driving can be adiabatically turned on without destroying the nomenclature of the states. In particular, the driving does not change the parity. A superposition of the ground and the first excited states is then not an eigenstate of the parity operator $P$. If the driving frequency is  brought close to resonance, $\omega_F\approx 2\omega_0$, so that the energy of the excited state in the RWA is extremely small, the superposition of the states has broken time-translation symmetry if measured on the time scale small compared to $\ep_1^{-1}$. This time is exponentially long for small $|(\omega_F/2)-\omega_0|$. The period doubling is thus topologically protected.

\section{Conclusions}

The results of this paper explicitly show that, in the quantum regime, subharmonics, including high-order ones, can be displayed already by the simplest driven quantum system, a qubit. The  proposed protocol is to drive the qubit by periodic pulses of a resonant field and/or by dc pulses that modulate the spacing between the qubit energy levels. The period of the pulses should be much longer than the reciprocal qubit transition frequency $\omega_0^{-1}$. Such driving relies on the driving conventionally used to perform qubit gate operations. The symmetry-broken states can be detected using conventional measurement protocols. Therefore an experiment on the symmetry breaking can be done with any qubits that have a high transition frequency.  

The ``price'' for the simplicity of the system is that, formally, the effect requires fine tuning of the parameters of the drive. However, given that qubits have a finite coherence time, one can establish periodicity of the dynamical variables of a qubit only with a limited precision. This limitation leads to a finite width of the parameter range within which the measured period is seen as a multiple of the drive period.

The other system considered in the paper is  a qubit chain with periodically modulated qubit coupling. The results refer to qubits with the transition frequency $\omega_0$  much higher than the coupling energy divided by $\hbar$, and to a resonant modulation with frequency $\omega_F$ close to $2\omega_0$. Such chain preserves the parity of the total number of  spin excitations. We found that, even for two spins, the quasienergies of states with opposite parity can differ by $\omega_F/2$, so that the supersposition of these states is periodic with period $4\pi/\omega_F$. 

A resonantly modulated qubit chain has a nontrivial time-translation symmetry and maps onto the Kitaev chain with the parameters controlled by the modulation. In a long chain, the state with a broken symmetry of time translation by $2\pi/\omega_F$ is topologically protected.  Speaking more broadly, such a qubit chain allows one to address, in a controlled experiment, several important problems of the Majorana physics. One of them is the effect of disorder \cite{Motrunich2001}. The site disorder can be emulated just by making the transition frequencies of the qubits slightly different. By making the coupling constants $J_{xx}, J_{yy}$ and the modulation amplitude $F$ site-dependent, one can emulate hopping and pairing disorder. If the system allows incorporating the $zz$-coupling, for example, if the Hamiltonian has the term $\sum_n\sigma_n^z\sigma_{n+1}^z$, one can also explore the effects of the fermion-fermion coupling. 

Quantum simulations with modulated qubit chains are invaluable in terms of studying the aforementioned effects.  Fluxonium qubits \cite{Manucharyan2009} may be advantageous, since modulation at frequency $\sim 2\omega_0$ will not lead to resonant transitions in such qubits. This will allow one to avoid leaving the computational subspace and the associated heating of the qubit system, which is of potential concern for transmon qubits, for example. A chain of $\sim 20$ qubits may already demonstrate the involved physics. On the other hand, a 20-qubit chain is presumably beyond what can be studied in the near future with classical computers, as it requires diagonalizing a $2^{20}\times 2^{20}$ Hamiltonian matrix, cf.~\cite{Borgonovi2016}.

I am grateful for the discussion to Vadim Smelyanskiy, Yu Chen, Vladimir Manucharyan, Pedram Roushan, and  Maxim Vavilov. This work was supported in part by the NSF, grants DMR-1708331 and DMR-1806473, and by the Google faculty research award.


%

\end{document}